\documentclass[
reprint,
aps,
prx,
amsmath, amssymb
]{revtex4-2}
\usepackage{graphicx, mathtools, xcolor, MnSymbol, placeins}
\usepackage{hyperref}
\usepackage[capitalize]{cleveref}
\usepackage[T1]{fontenc}

\newcommand{\bk}[0]{\mathbf{k}}

\newcommand{\bq}[0]{{\mathbf{q}}}

\newcommand{\lr}[1]{\left( #1 \right)}
\newcommand{\lrb}[1]{\left[ #1 \right]}

\begin{document}
\title{Tunable Superconductivity Mediated by Heavy-Electron Plasmons:\\
Band-Structure and Quantum-Geometric Engineering}
\author{Sang Hyun Park}
\affiliation{Department of Physics, The University of Texas at Austin, Austin, TX 78712, USA}
\author{Junyeong Ahn}
\email{junyeong.ahn@austin.utexas.edu}
\affiliation{Department of Physics, The University of Texas at Austin, Austin, TX 78712, USA}
\date{\today}

\begin{abstract}
Conventional superconductivity derives its pairing glue from lattice vibrations, tying its characteristic scales to chemistry and atomic masses. Plasmons---the collective oscillations of electrons---can instead be reshaped through electronic structure engineering, but the principles governing optimal plasmon-mediated pairing remain unclear. Here, we establish such principles for two-carrier systems in which heavy-electron plasmons mediate the pairing of light electrons. Within the random-phase approximation and Eliashberg theory, we calculate the optimal $T_c$ of minimal metallic models and show that it is controlled by a competition between the plasmon energy scale and retardation-driven suppression of the repulsion, yielding optimal carrier densities and band masses. While the plasmon channel alone reaches only $T_c\sim$ 0.1 K, a moderate phonon attraction cooperates with it, boosting $T_c$ by two orders of magnitude to above 20 K.  However, the band flattening needed for slow metallic plasmons also favors the development of competing orders. We therefore consider an insulating system in which coherent interband transitions between flat bands generate gapped interband plasmons without free carriers. The heavy-band quantum metric governs the dispersion and electron-plasmon pairing strength of the interband plasmon, while the quantum geometry of the light band suppresses static screening and enhances the net attraction. Because layer separation rapidly weakens pairing, we propose systems with coexisting light and heavy electrons living in different mirror-symmetry sectors of the same layer as promising platforms. Our results establish a new role for flat-band systems in superconductivity: rather than hosting the paired electrons themselves, they can serve as a tunable pairing mediator whose collective charge excitations set the superconducting energy scale beyond their narrow bandwidth.
\end{abstract}
\maketitle

\begin{figure}
    \centering
    \includegraphics{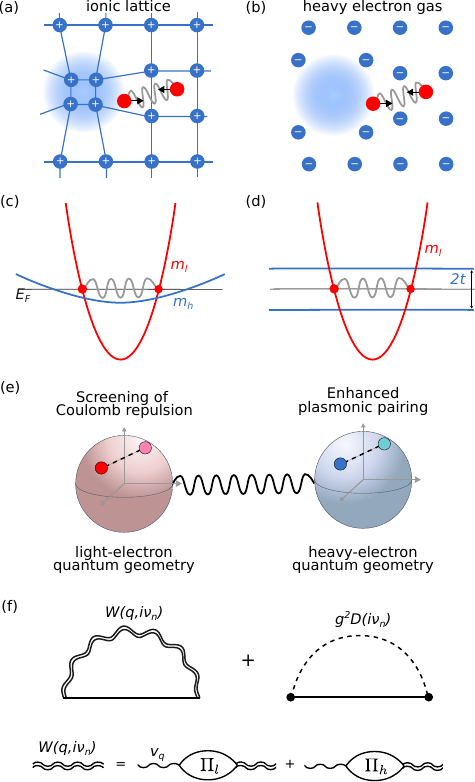}
    \caption{Tunable superconductivity mediated by heavy-electron plasmons. (a) A schematic representation of electron pairing due to a phonon-mediated attractive interaction. The ionic lattice which supports the phonons are shown as blue circles while the pairing electrons are shown in red. (b) A schematic representation of electron pairing due to a plasmon-mediated attractive interaction. The heavy electron gas which supports the plasmons are shown as blue circles. (c) The band structure for a two-carrier model in which the plasmons of a metallic heavy electron band (blue) can mediate superconductivity in a light electron band (red). (e) Schematic representation of the quantum geometry of the light and heavy electrons. (f) Feynman diagrams showing the one-loop electron self-energy with contributions from the screened Coulomb interaction and a phonon. The screened Coulomb interaction is calculated in the random phase approximation including contributions from both the light and heavy electrons.}
    \label{fig:schematic}
\end{figure}

\section{Introduction}
Phonons, the collective vibrations of the atomic lattice, play a central role in conventional superconductors by producing a retarded attractive interaction between electrons. Within Bardeen-Cooper-Schrieffer (BCS) theory~\cite{Bardeen1957} and Eliashberg theory~\cite{Eliashberg1960}, the superconducting transition temperature $T_c$ is closely tied to the characteristic phonon energy scale. This connection is reflected in the isotope effect, where $T_c\sim M^{-\beta}$, with $M$ the atomic mass and $\beta$ approaching the ideal value of $0.5$ in many conventional superconductors~\cite{Carbotte1990}. This observation has motivated the search for higher-$T_c$ superconductivity in materials composed of lighter elements, with hydrogen representing the ultimate limit~\cite{Ashcroft1968}. Indeed, conventional high-temperature superconductivity has been observed in hydrides ~\cite{Drozdov2015, Drozdov2019, Somayazulu2019, Flores-Livas2020}, but only under very high pressures. More generally, phonon-mediated superconductivity is constrained by the limited tunability of lattice vibrations and by the need to identify suitable materials. A recent computational study~\cite{Gao2025} surveyed the electron-phonon interaction and superconductivity in more than 20,000 metals and concluded that the experimental realization of high-$T_c$ materials at ambient pressure is extremely unlikely.

These limitations motivate the search for alternative collective modes that can mediate a retarded attraction. Plasmons, the collective charge oscillations of electrons, provide a natural possibility, as they can also generate a retarded interaction between electrons and mediate superconductivity~\cite{Takada1978, Akashi2013, Ruhman2016, Fatemi2018, Sharma2020, intVeld2026}. A key advantage of plasmons is that their properties are determined by the electronic band structure and can therefore be widely engineered in two-dimensional materials through electrostatic gating~\cite{Fei2012}, twisting~\cite{Sunku2018, Hesp2021, Huang2022}, and substrate engineering~\cite{Xiong2019, Xiong2021}. This offers a possible route toward tunable superconductivity beyond the material constraints of phonon-mediated pairing. However, plasmons formed by the superconducting electrons themselves typically have energies comparable to or larger than the Fermi energy and therefore cannot efficiently provide the retarded attractive interaction required for pairing~\cite{Grabowski1984, Grimaldi1995}.

In this work, we explore superconductivity in a two-carrier system, where plasmons of heavy electrons generate a retarded interaction that mediates superconducting pairing among light electrons. We first consider light-electron pairing mediated by intraband plasmons in a metallic heavy-electron system and study the dependence of $T_c$ on carrier density and band dispersion by numerically solving the Eliashberg equations. We find that flattening the heavy-electron band plays a critical role in enhancing the optimal $T_c$ by making retardation effects stronger. While promising, strong flattening of a metallic heavy-electron band can also make the system susceptible to competing ordered phases that may preempt plasmon-mediated superconductivity~\cite{Cao2018a, Bultinck2020, Chen2020b, Regan2020, He2021}. To avoid such non-superconducting phases and stabilize the heavy-electron plasmon, we further examine interband plasmons in insulating systems with a pair of flat bands. In this case, collective modes arise from coherent interband transitions rather than intraband motion of free carriers. We find that the quantum geometry of the heavy-electron bands dictates both the interband plasmon dispersion and the electron-plasmon coupling, giving rise to a strong dependence of $T_c$ on the heavy-electron quantum geometry.
Finally, we turn our attention to the quantum geometry of the light electrons which form the superconducting Cooper pairs. It has recently been shown that non-trivial quantum geometry modifies the static polarization of a two-dimensional electron gas which leads to an enhanced Kohn-Luttinger superconductivity~\cite{Shavit2025, Jahin2026}. For the plasmon-mediated superconducting case, we find that the non-trivial quantum geometry enhances the net pairing strength and the transition temperature in the isotropic channel.

We begin by outlining the conceptual similarities of the phonon- and plasmon-mediated superconducting mechanisms. When an electron interacts with the underlying ionic lattice, it can excite a phonon, which in turn induces local variations in the charge polarization that can attract a second electron. The response time of the induced charge polarization of the ionic lattice, i.e. the phonon, is slow compared to that of the electron due to the large difference in masses, and thus results in a retarded attraction between electrons. This process is schematically shown in \cref{fig:schematic}(a). From this qualitative picture, we can see that the existence of a charged collective mode with a response time slower than that of the pairing electrons is the key ingredient for superconductivity. The attractive nature of the retarded interaction does not depend on the sign of the ionic charge. Therefore, we can also expect plasmons, which are the collective excitations of an electron gas, to be able to generate an attractive interaction between the electrons, as shown in \cref{fig:schematic}(b).

\begin{figure*}
    \centering
    \includegraphics{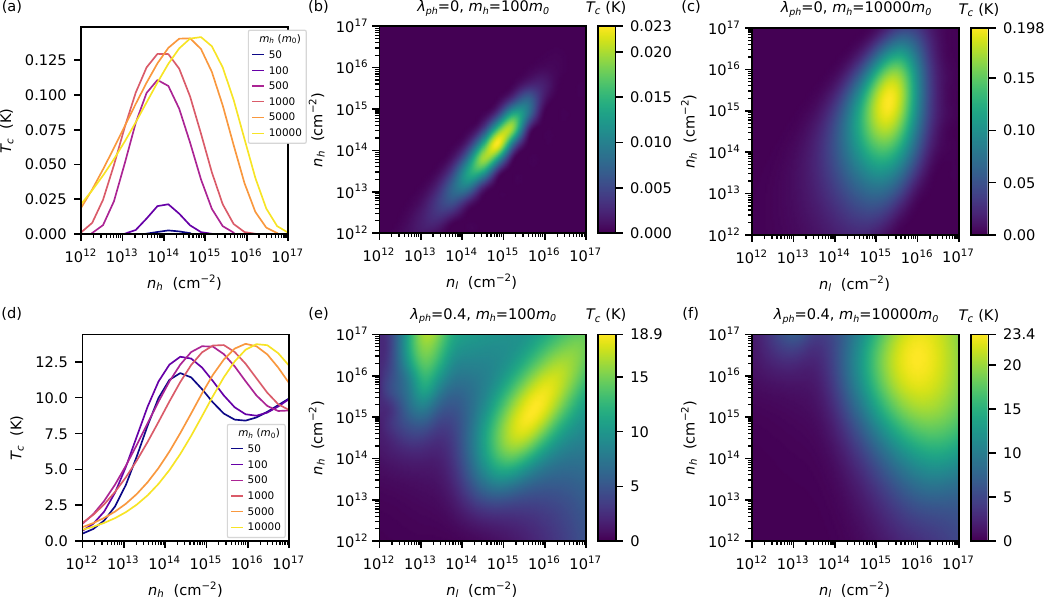}
    \caption{Calculations of the $T_c$ using the linearized isotropic Eliashberg equations for the setup shown in \cref{fig:schematic}(c). (a-c) Plasmon-only results as a function of the light and heavy electron carrier density and effective mass. (d-f) Results that include a contribution from the phonon with coupling strength $\lambda_{\rm ph}=0.4$. For (a),(d) the light electron carrier density is fixed to $n_l=5\times10^{14}\mathrm{cm}^{-2}$. The light electron mass is fixed to $m_l=m_0$.}
    \label{fig:intraband}
\end{figure*}

However, while any conduction electron supports plasmon excitations, they are typically at an energy larger than the Fermi energy and therefore are not able to provide a retarded interaction. This limitation can be overcome when two distinct electronic quasiparticle species have well-separated mass scales, such that the collective mode of one of the species is much slower than the electrons that form Cooper pairs. The simplest setting that satisfies this condition consists of two parabolic bands with markedly different effective masses [\cref{fig:schematic}(c)]. The light electrons with mass $m_l$ will form the Cooper pairs while the heavy electrons with mass $m_h$ will provide the plasmonic pairing glue. This setting is discussed in \cref{sec:intraband}. A second setup that can also provide the separation in energy scales between the pairing electrons and the plasmons is by considering interband plasmons in insulating flat bands [\cref{fig:schematic}(d)]. Here, the energy scale of the plasmonic pairing glue is determined by the gap size $2t$ and the Brillouin zone (BZ) averaged quantum metric. This setup is discussed in \cref{sec:interband}. Finally, we discuss the effects of quantum geometry in the pairing light electron bands in \cref{sec:screening}. The interband plasmon pairing of the heavy electron bands and the screening from the light electrons both originate from quantum geometry [\cref{fig:schematic}(e)].

Throughout this work, we focus on two-dimensional systems, motivated primarily by layered and moiré materials. The conceptual framework and Eliashberg treatment can also be applied to three-dimensional systems, with appropriate modifications to the Coulomb interaction, electronic polarization, and plasmon dispersion.

\section{Intraband plasmons in metallic heavy electron bands}
\label{sec:intraband}

In this section, we consider the setting shown in \cref{fig:schematic}(c) consisting of two parabolic bands with different effective masses. The plasmon-mediated pairing can be modeled by the screened Coulomb interaction within the random phase approximation (RPA) [\cref{fig:schematic}(f)]. Including contributions from both the light and heavy electrons, we may write
\begin{equation}\label{eq:Wfull}
    W(q,i\nu_n)=\frac{v_q}{1-v_q\lrb{\Pi_l(q,i\nu_n) + \Pi_h(q,i\nu_n)}}
\end{equation}
where $v_q=e^2/2\epsilon_{{\rm env}} q$ is the bare Coulomb interaction and $\epsilon_{{\rm env}}$ is the effective dielectric constant due to high-energy bands not included in the model and the surrounding environment. The bare polarization functions of each carrier type $i=l,h$ is 
\begin{equation}
    \Pi_i(q,i\nu_n)=2\int_\bk\frac{f_{i,\bk} - f_{i,\bk+\bq}}{i\hbar\nu_n+E_{i,\bk}-E_{i,\bk+\bq}}
\end{equation}
where $\int_{\bk}=\int d^2\bk/(2\pi)^2$, $f_{i,\bk}$ is the Fermi-Dirac distribution, $E_{i,\bk}=\hbar^2k^2/2m_i-E_{F,i}$ is the energy dispersion, $E_{F,i}$ is the Fermi energy, and $\nu_n=2n\pi k_BT/\hbar$ are the bosonic Matsubara frequencies. The electron self-energy is then approximated using the one-loop diagram with the screened Coulomb interaction as shown in \cref{fig:schematic}(f) ~\cite{Margine2013}.

The characteristic frequency of both the light and heavy electron gases is determined by the plasmon dispersion $\omega_{p}(q)$. In the limit $\omega\gg qv_F$, the plasmon dispersion of the two-dimensional electron gas is given by $\omega_p^2(q)=ne^2q/2m\epsilon_{\rm env}$ where $n$ is the carrier density. At $q=k_F$, the plasmon energy is $\hbar\omega_p(k_F)=2^{3/4}\sqrt{r_s}E_F$ where $r_s=1/a_B^*\sqrt{\pi n}$, and $a_{B}^*=4\pi\epsilon_{\rm env}\hbar^2/me^2$ is the effective Bohr radius. As noted before, we find that the plasmon energy of the light electrons is typically larger than the Fermi energy $E_{F,l}$. For a retarded interaction between the light electrons, we are interested in energies that are smaller than $E_{F,l}$ and may approximate the light electron polarization using its static limit $\Pi_l(q,0)=-2N_{F,l}$ where $N_{F,l}=m_l/2\pi\hbar^2$ is the density of states per spin at the Fermi energy. Using this approximation, the screened interaction can be written as
\begin{equation}\label{eq:W}
    W(q,i\nu_n)=\tilde v_q\lrb{1 + \frac{\tilde v_q\Pi_h(q,i\nu_n)}{1-\tilde v_q\Pi_h(q,i\nu_n)}}
\end{equation}
where $\tilde v_q=e^2/2\epsilon_{\rm env}(q+\kappa)$ is the static screened Coulomb interaction and $\kappa=e^2N_{F,l}/\epsilon_{\rm env}=2/a_{B}^*$ is the Thomas-Fermi screening wave vector of the light electrons. The first term in \cref{eq:W} gives the static repulsion between light electrons while the second term gives the plasmon attraction. The plasmon attraction can be seen clearly by approximating $\Pi_h\approx -n_hq^2/m_h\nu_n^2$ and writing the screened interaction as
\begin{equation}\label{eq:W2}
    W(q,i\nu_n)=\tilde v_q\lrb{1 - \frac{\Omega_{p,h}^2(q)}{\nu_n^2+\Omega_{p,h}^2(q)}}
\end{equation}
where the screened heavy-electron plasmon dispersion is $\Omega_{p,h}^2(q)=\omega^2_{p,h}q/(q+\kappa)$.

To calculate the superconducting $T_c$ of the system, we solve the linearized isotropic Eliashberg equations ~\cite{Margine2013}
\begin{subequations}
    \begin{align}
        Z(i\omega_n)&=1+\frac{\pi k_B T_c}{\hbar\omega_n}\sum_m \mathrm{sgn}(\omega_m)\lambda(i\omega_n-i\omega_m) \\
        \Delta(i\omega_n)&=\frac{\pi k_BT_c}{Z(i\omega_n)}\sum_m\frac{\Delta(i\omega_m)}{\hbar|\omega_m|}\lrb{\lambda(i\omega_n-i\omega_m) - \mu}
    \end{align}
\end{subequations}
where $\omega_n=(2n+1)\pi k_BT/\hbar$ are the fermionic Matsubara frequencies. For all calculations, we choose a cutoff of $E_{F,l}/\hbar$ for the Matsubara frequencies. The dimensionless static Coulomb repulsion $\mu$ and dynamic plasmonic attraction $\lambda(i\nu_n)$ are defined as
\begin{equation}
    \mu=N_{F,l}\langle \tilde v_q\rangle_{\mathrm{FS}},\quad \lambda(i\nu_n)=\mu - N_{F,l}\langle W(q,i\nu_n)\rangle_{\mathrm{FS}}
\end{equation}
and the Fermi surface average is defined as $\langle f(q)\rangle_{\mathrm{FS}}=\frac{1}{2\pi}\int_{-\pi}^\pi d\theta f(k_F\sqrt{2-2\cos\theta})$. We use \cref{eq:W} for the screened interaction $W(q,i\nu_n)$ in our Eliashberg calculations. Finally, we will also consider the effect of including an additional phonon-mediated attraction by adding a term $\lambda_{\rm ph}\omega_{\rm{ph}}^2/(\nu_n^2+\omega_{\rm{ph}}^2)$ to the attractive $\lambda(i\nu_n)$. The total electron self-energy is shown in \cref{fig:schematic}(f).

We first examine the dependence of $T_c$ on the heavy electron carrier density $n_h$ and mass $m_h$ in the absence of any phonon contributions ($\lambda_{\rm ph}=0$). The results shown in \cref{fig:intraband}(a) exhibit a dome-like behavior as a function of the carrier density $n_h$ for all heavy electron masses $m_h$. This result can be understood by considering the McMillan-Allen-Dynes equation for $T_c$~\cite{McMillan1968,Allen1975},
\begin{equation}\label{eq:mcmillan}
    T_c=\frac{\hbar\omega_{\mathrm{log}}}{1.2k_B}\exp\lr{-1.04\frac{1+\lambda}{\lambda-\mu^*(1+0.62\lambda)}}
\end{equation}
where $\omega_{\mathrm{log}}$ is the log-averaged plasmon frequency, $\lambda=\lambda(0)$ is the static attraction, $\mu^*=\mu/(1+\mu\log E_F/\hbar\omega_c)$ is the Anderson-Morel Coulomb potential~\cite{Morel1962}, and $\omega_c$ is the plasmon dispersion bandwidth. The renormalized Coulomb potential $\mu^*$ is what encodes the retarded nature of the plasmon attraction. Note that the plasmon frequency enters into two separate parts of \cref{eq:mcmillan}. The overall prefactor $\omega_{\mathrm{log}}$ determines the energy scale of $T_c$ and is responsible for the initial enhancement of $T_c$ as $n_h$ increases. On the other hand, the reduction of the Coulomb potential $\mu^*$ is harmed by increasing the plasmon frequency as it has a denominator that depends on the retardation condition $E_F/\hbar\omega_c\gg1$. The competition between these two effects is what is responsible for the dome-like behavior with respect to $n_h$. 

We also see that flattening of the heavy electron bands leads to an enhancement of the optimal $T_c$ [\cref{fig:intraband}(a)]. At a fixed heavy electron carrier density $n_h$, we observe a dome-like behavior with respect to the heavy electron mass $m_h$. Since the plasmon frequency is inversely proportional to the electron mass $\omega_p\sim 1/\sqrt{m}$, heavier electron masses give lower plasmon frequencies which are then able to effectively suppress the Coulomb repulsion from its bare value $\mu$ to the renormalized $\mu^*$. 

In \cref{fig:intraband}(b) and (c), we calculate the dependence of $T_c$ on both the heavy and light electron carrier densities and find that the dome-like behavior is also observed as a function of the light electron carrier density $n_l$. The initial increase in $T_c$ can once again be tracked back to the retardation condition $E_F/\hbar\omega_c\gg1$ since increasing the light electron carrier density increases $E_F$. To understand the suppression of $T_c$ at higher $n_l$, we use the approximate form of $W(q,i\nu_n)$ given in \cref{eq:W2} to find the analytic form of the static attraction $\lambda$ to be
\begin{equation}\label{eq:static_lda}
\lambda =
\begin{cases}
 \frac{1}{\pi\sqrt{1-1/x^2}}\arctan\lr{\sqrt{x^2-1}}    & x>1 \\
 \frac{1}{\pi\sqrt{1/x^2-1}}\ln\lr{1/x+\sqrt{1/x^2-1}} & 0<x<1
\end{cases}
\end{equation}
where $x=\kappa/2k_F$. This function is a monotonically increasing function of $x$ with an asymptotic value of 0.5 at $x\rightarrow \infty$. As the light-electron carrier density $n_l$ increases, $k_F$ increases while $\kappa$ remains constant, leading the decrease in $x$. Since $\lambda(x)$ function is a monotonically increasing function of $x$, we can expect the pairing strength to decrease as $n_l$ increases.

The effect of including phonon attraction is shown in \cref{fig:intraband}(d-f). We find that while the phonon and plasmon are individually only capable of reaching a $T_c$ on the order of 0.1 K, the collaboration between both mechanisms can boost the $T_c$ by roughly two orders of magnitude to reach 20 K. Otherwise, we find that the qualitative dependence of $T_c$ on both the heavy and light electron parameters to be unchanged when the phonon is included.

\begin{figure*}[t]
    \centering
    \includegraphics{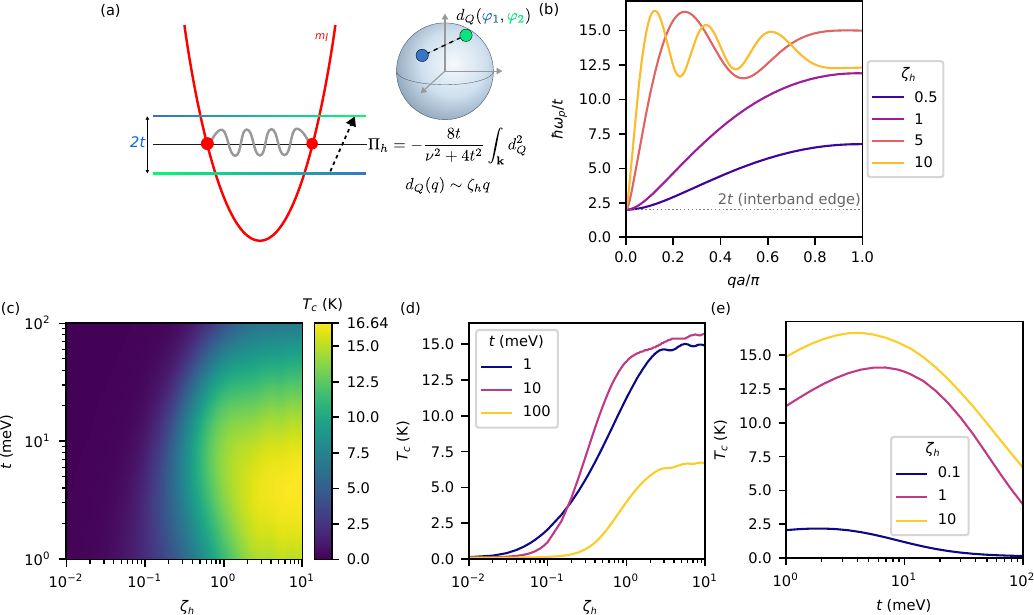}
    \caption{Quantum-geometric control of interband-plasmon-mediated superconductivity in gapped flat bands. (a) Schematic of the setup: a pair of flat bands separated by an energy $2t$ supports a pairing interaction mediated by gapped interband plasmons (wavy line). The interband plasmons arise from coherent interband transitions (dashed arrow) rather than free-carrier motion. The associated interband polarization is controlled by the BZ averaged quantum distance $d_Q$, and the light electrons (red, mass $m_l$) form the Cooper pairs. The inset is a schematic representation of the quantum distance between two states on the Bloch sqhere. (b) Interband plasmon dispersion $\hbar\omega_p/t$ versus $qa/\pi$ for several values of the heavy-electron quantum-metric parameter $\zeta_h$. (c) Superconducting transition temperature $T_c$ as a function of $\zeta_h$ and gap size $t$. Light electron parameters are fixed to $m_l=m_0,\ n_l=5\times 10^{14}\mathrm{cm}^{-2}$ and a phonon term with $\lambda_{\rm ph}=0.4$ is included. (d,e) Slices of (c) along the $\zeta_h$ and $t$ axes.}
    \label{fig:interband}
\end{figure*}

\section{Interband plasmons in insulating flat bands}
\label{sec:interband}
In this section we consider interband plasmons in insulating flat bands as the pairing glue for plasmon-mediated superconductivity [\cref{fig:interband}(a)]. Such a setup can be expected to be more robust against developing competing ordered states that may preempt plasmon-mediated superconductivity. We specifically consider the tunable metric model that was introduced in ~\cite{Hofmann2022, Hofmann2023} with the Hamiltonian
\begin{equation}\label{eq:tm_model}    h(\bk)=t\lr{\tau_x\sin\alpha_\bk+\sigma_z\tau_y\cos\alpha_\bk}-\mu
\end{equation}
where $\tau_i$ $(\sigma_i)$ are Pauli matrices that act on the orbital (spin) degrees of freedom and $\alpha_\bk=\zeta(\cos k_xa+\cos k_ya)$. This Hamiltonian has flat bands separated by $2t$ and a Brillouin zone (BZ) integrated quantum metric given by
\begin{align}
    G_{ij}\equiv \int_{\bf k} g_{ij}=\delta_{ij}\zeta^2/8
\end{align}
where $i,j$ are spatial coordinate indices. 

When the Fermi energy is in the gap between the flat bands, the polarization can be written as
\begin{equation}\label{eq:interband_polarization}
    \Pi(\bq,i\nu_n)=-\frac{4E_g}{(\hbar\nu_n)^2+E_g^2}
    \int_{\bk} d_Q^2(\varphi_{\bk,-},\varphi_{\bk+\bq,-})
\end{equation}
where $E_g$ is the constant energy gap, $d_Q(\varphi_{1},\varphi_{2})=\sqrt{1-|\langle\varphi_{1}|\varphi_{2}\rangle|^2}$ is the Hilbert-Schmidt quantum distance~\cite{Provost1980} and $\varphi_{\bk,\pm}$ is the eigenstate of $\pm t$ band. Equation~\eqref{eq:interband_polarization} applies to any two-band systems with a constant energy gap. In our model, $E_g=2t$. For small $q$ we can approximate the average quantum distance squared as
\begin{align}
    d^2_Q(\bq)\equiv \int_{\bk}d_Q^2(\varphi_{\bk,-},\varphi_{\bk+\bq,-})\approx \sum_{ij}q_iq_jG_{ij}=\frac{1}{8}\zeta^2q^2.
\end{align}
From the condition $1+\kappa/q-v_q\Pi(q,\omega)=0$ we find the screened interband plasmon dispersion to be 
\begin{align}\label{eq:screened_pl}
    \hbar\Omega_p(q)&=2t\sqrt{1+\frac{2v_q}{t}\frac{q}{q+\kappa}d^2_Q(q)}\approx2t\sqrt{1+\frac{e^2\zeta^2}{8t\epsilon_{\rm env}}\frac{q^2}{q+\kappa}}.
\end{align}
The interband plasmon dispersion as a function of the parameter $\zeta$, which controls the quantum metric, is shown in \cref{fig:interband}(b). Note that the interband origin of the plasmon results in a gapped spectrum starting at $2t$. Also, as expected from \cref{eq:screened_pl}, we can see that the plasmon increases in energy as a function of $\zeta$. Interestingly, the plasmon energy does not increase indefinitely, but reaches a maximum energy and exhibits an oscillatory behavior. Such behavior originates from oscillations in the average quantum distance $d_Q(q)$ and have also been identified for the static polarization~\cite{Shavit2025}.

\begin{figure*}
    \centering
    \includegraphics{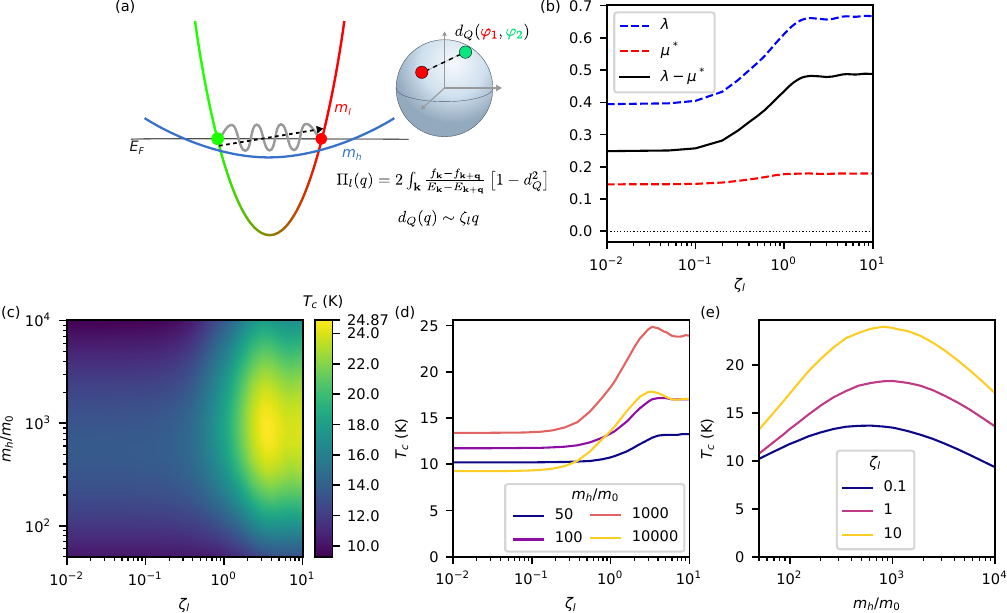}
    \caption{Enhancement of heavy-electron-plasmon-mediated superconductivity by light-electron quantum geometry. (a) Schematic of the setup: the light (pairing) electrons occupy a parabolic band (mass $m_l$) endowed with non-trivial quantum geometry, obtained from the tunable metric model of \cref{eq:tm_model} plus an added trivial square-lattice dispersion, while the heavy electrons (mass $m_h$) supply the plasmon pairing glue. The non-trivial geometry suppresses the static screening (dashed arrow) from the light electrons. (b) Dimensionless plasmonic attraction $\lambda$, renormalized Coulomb repulsion $\mu^*$, and net pairing strength $\lambda - \mu^*$ as functions of the light-electron quantum-metric parameter $\zeta_l$. (c) Superconducting transition temperature $T_c$ versus $\zeta_l$ and $m_h$. Carrier densities are fixed to $n_l=5\times10^{14}\mathrm{cm}^{-2},\ n_h=10^{15}\mathrm{cm}^{-2}$, and a phonon term with $\lambda_{\rm ph}=0.4$ is included. (d,e) Slices of (c) along the $\zeta_l$ and $m_h$ axes.}
    \label{fig:screening}
\end{figure*}

To calculate $T_c$ we use the interband polarization \cref{eq:interband_polarization} as $\Pi_h$ in the screened interaction \cref{eq:W} to find
\begin{equation}
    W(q,i\nu_n)=\tilde v_q\lrb{1-\frac{\Omega_p^2-4t^2/\hbar^2}{\nu_n^2+\Omega_p^2}}.
\end{equation}
Solutions of the linearized Eliashberg equation as a function of $\zeta_h$ and $t$ are shown in \cref{fig:interband}(c-e). We have included the subscript $h$ in $\zeta_h$ to clarify that we are tuning the quantum geometry of the heavy-electron bands. For all values of $t$, we find that the $T_c$ initially increases rapidly with $\zeta_h$ after which it plateaus. This behavior can be traced back to the existence of a maximum plasmon energy at large $\zeta_h$. 

\section{Enhanced pairing from quantum geometric screening}
\label{sec:screening}
We now turn our attention to the screening $\kappa$. From \cref{eq:W}, we can see that the static screened Coulomb interaction $\tilde v_q$ sets the overall magnitude of both the repulsion $\mu$ and the attraction $\lambda$. Hence, it can be expected that the screening vector $\kappa$ has a significant impact on $T_c$. In general, the screening vector is given by $\kappa=-e^2\Pi_l(q)/2\epsilon_{\rm env}$ where $\Pi_l(q)$ is the static polarization of the light electrons~\cite{Shavit2025}
\begin{equation}
    \Pi_l(q)=2\int_\bk\frac{f_\bk-f_{\bk+\bq}}{E_\bk-E_{\bk+\bq}}\lrb{1-d^2_Q(\varphi_\bk,\varphi_{\bk+\bq})}.
\end{equation}
For a two-dimensional electron gas, it is well known that the polarization is constant up to $2k_F$, which gives a constant screening vector $\kappa=e^2N_F/\epsilon_{\rm env}$. However, if the electron bands have a non-trivial quantum geometry, the screening vector will be suppressed below $e^2N_F/\epsilon_{\rm env}$ and acquire a $q$ dependence. 

Since the screening is due to the light electrons, we will need to add a non-trivial quantum geometry to the superconducting light electron bands as schematically shown in \cref{fig:screening}(a). To do this, we take the tunable metric model Hamiltonian \cref{eq:tm_model} and add a trivial square lattice dispersion $2t'(\cos k_xa+\cos k_ya)\tau_0\sigma_0$ term. In the limit $t\gg t'$ and near the band edge, we get a parabolic band model with non-trivial quantum geometry. 

In \cref{fig:screening}(b), we first examine the dependence of $\lambda$ and $\mu^*$ on the parameter $\zeta_l$ which controls the light-electron quantum metric. We find that increasing $\zeta_l$ results in an enhancement of both the attractive $\lambda$ and repulsive $\mu^*$. However, the enhancement in the attraction outweighs the enhancement in the repulsion and boosts the net attraction $\lambda-\mu^*$. This increase in net attraction is clearly observed in the Eliashberg calculations presented in \cref{fig:screening}(c-e).

\section{Effect of layer separation and dielectric environment}

\begin{figure}
    \centering
    \includegraphics{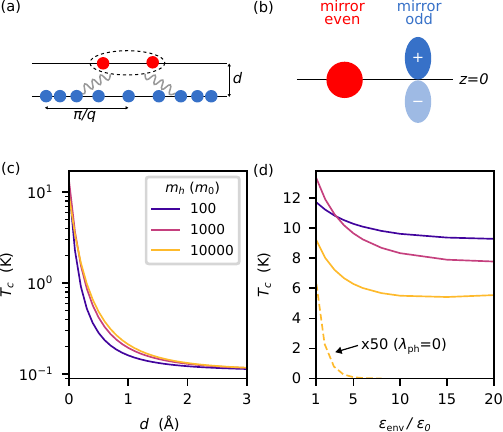}
    \caption{Dependence on layer separation and screening from the dielectric environment. (a) Schematic representation of layer separation between the light electrons (red) and the heavy electrons (blue). The heavy electrons support plasmons of wavelength $2\pi/q$. (b) Schematic representation of a mirror symmetric system with coexisting mirror-even and mirror-odd orbitals with center of mass in the $z=0$ plane. Dependence of $T_c$ on the (c) layer separation $d$ and (d) dielectric constant of the environment $\epsilon_{\mathrm{env}}$. Carrier densities are fixed to $n_l=5\times10^{14}\mathrm{cm}^{-2},\ n_h=10^{15}\mathrm{cm}^{-2}$ and the phonon term is fixed to $\lambda_{\rm{ph}}=0.4$. The dashed yellow line in (d) is the $\lambda_{\rm{ph}}=0$ result for $m_h=10000m_0$ scaled by a factor of 50 for visibility.}
    \label{fig:layersep_eps}
\end{figure}

Before concluding, we analyze how separating the heavy and light electrons into different layers, as well as the choice of dielectric environment, affects the system. The layer separation can be incorporated by modifying the screened potential \cref{eq:Wfull} to
\begin{equation}
    W(q,i\nu_n)=v_q\frac{1-v_q\Pi_h\lr{1-e^{-2qd}}}{1-v_q\Pi_l-v_q\Pi_h+v_q^2\Pi_l\Pi_h(1-e^{-2qd})}
\end{equation}
where $d$ is the layer separation [\cref{fig:layersep_eps}(a)]. Note that we have dropped the arguments $q,i\nu_n$ for the polarizations $\Pi_i$ on the right-hand side for notational convenience. The dependence of $T_c$ on layer separation for the intraband plasmon mechanism [\cref{sec:intraband}] is shown in \cref{fig:layersep_eps}(c). We find that the $T_c$ rapidly decays with respect to the separation such that at $3$\AA, only the $T_c$ from the phonon term remains. It is thus essential that the layer separation is minimized, or if possible, have the light and heavy electrons live in the same layer. Mirror symmetric systems in which the light and heavy electrons live in distinct sectors provide a promising platform for minimizing the layer separation and maximizing the $T_c$ [\cref{fig:layersep_eps}(b)]. In this case, two electron sectors do not hybridize at the single-particle level but can interact through Coulomb interactions. A further discussion of materials is provided in \cref{sec:materials}.

The $T_c$ dependence on the dielectric environment is shown in \cref{fig:layersep_eps}(d). We find that the $T_c$ initially decreases as a function of $\epsilon_\mathrm{env}$ and eventually plateaus for large $\epsilon_\mathrm{env}$. This behavior can be understood using the McMillan parameters $\lambda$ and $\mu^*$. An increasing $\epsilon_\mathrm{env}$ will decrease the size of both $\lambda$ and $\mu^*$. The net plasmonic pairing strength $\lambda-\mu^*$ also decreases. In the absence of the phonon term, we find that this results in a complete suppression of $T_c$, as shown by the dashed line in \cref{fig:layersep_eps}(d). In the presence of the constant phonon term $\lambda_{ph}$, the reduced repulsion $\mu^*$ will enhance the $T_c$ originating from the phonon, which determines the $T_c$ at large $\epsilon_{\mathrm{env}}$. Note that the bare phonon $T_c$ at $\epsilon_{\mathrm{env}}=\epsilon_0$ is $\sim 0.1$K, as observed from \cref{fig:layersep_eps}(c).

\section{Discussion}

Our results highlight the interplay between band structure and quantum geometry in engineering superconductivity mediated by heavy-electron plasmons. We now place these results in the context of previous studies, discuss the prospects for enhancing the transition temperature, and identify candidate material platforms in which the proposed mechanisms may be realized.

\subsection{Relation to existing work}
We first situate our results within previous studies of superconductivity in two-carrier systems and of quantum-geometric
effects on superconductivity. Coulombic engineering of superconductivity in layered materials has been studied in ~\cite{Fatemi2018, intVeld2026}, where it was proposed that superconductivity can be controlled through intraband plasmons in a remote metallic screening layer.
Our results in \cref{sec:intraband} follow a similar path and examine the full parameter space for both the light and heavy electron layers. The results in \cref{sec:interband} and \cref{sec:screening} go further and examine the effects of quantum geometry on plasmon-induced superconductivity. An electron liquid-hole crystal bilayer was studied in ~\cite{Nashabeh2026} in which superconductivity induced by the plasmon modes of a Wigner crystal layer was proposed. The plasmon modes considered in our work are distinct from the Wigner crystal collective modes as they are either in the electron liquid regime [\cref{sec:intraband}, \cref{sec:screening}] or come from interband transitions [\cref{sec:interband}]. Two-band systems where the heavy-mass band edge is slightly above the Fermi energy have also been expected to show enhanced superconductivity due to pair scattering between the heavy and light bands~\cite{Ochi2022, Tajima2024}. This mechanism does not involve plasmons and is distinct from our work.

Quantum geometry has now been established as an essential ingredient for understanding flat-band superconductivity. Early works demonstrated that the superfluid weight of a flat band is bounded below by the BZ integral of the quantum metric~\cite{Peotta2015, Liang2017}. Specifically for twisted bilayer graphene, it was shown that the topology of the flat bands presents a lower bound for the superfluid weight~\cite{Xie2020}. The model \cref{eq:tm_model} was introduced in Refs.~\cite{Hofmann2022, Hofmann2023} as a topologically trivial model with continuously tunable quantum metric to study flat band superconductivity. The above mentioned works focused on the quantum geometry while treating the pairing interaction strength as a separate parameter. Recent works have shown further that quantum geometry can also directly enter the pairing interaction and enhance Kohn-Luttinger superconductivity~\cite{Shavit2025, Jahin2026}. Our results in \cref{sec:interband} and \cref{sec:screening} also consider the effect of quantum geometry on the pairing interaction, but focus on the dynamic isotropic pairing channel, in contrast to the works on Kohn-Luttinger superconductivity which focus on higher angular momentum channels in the static limit.

\subsection{Prospects for higher transition temperature}
We now comment on the prospect of reaching a higher $T_c$ with the plasmon-mediated mechanism. In conventional superconductors, the transition temperature is set by the Debye frequency, as exemplified by the isotope effect. In the two parabolic band system [\cref{fig:schematic}c], the plasmon frequency can be increased either by increasing the heavy electron carrier density $n_h$ or by making the heavy band more dispersive. This would naively suggest that raising the plasmon frequency should also raise $T_c$. Our results in \cref{fig:intraband} show that this is not always true. A high $T_c$ requires a balance between the plasmon energy scale and the retardation condition, rather than maximization of the plasmon frequency alone. However, even at the optimal point, the plasmon mechanism alone is limited to $T_c\sim 0.1$K. 

The pairing strength $\lambda$ for the plasmon mechanism is set by the screened interaction strength $e^2/2\epsilon_{\rm env}(q+\kappa)$. The limited $T_c$ observed for the plasmon-only case is thus tied to the screening vector $\kappa$ of the light electrons. For light electron bands with trivial quantum geometry, we have shown in \cref{eq:static_lda} that $\lambda\leq 0.5$, which is the fundametal reason behind the limited $T_c$ observed for the plasmon-only case. We thus demonstrated in \cref{sec:screening} that reduction of $\kappa$ by quantum geometry leads to an enhancement of the pairing strength $\lambda$ and ultimately the $T_c$. While our current results are model specific, the general mechanism of reducing the screening vector is applicable to any system with non-trivial quantum geometry.

One strategy for further enhancing $T_c$ is by considering the collaborative effect of the plasmon with other collective modes, which can enhance the total $\lambda$. For example, results in \cref{fig:intraband}(d-f) show that the addition of a phonon can enable the system to reach $T_c\sim20$K with a moderate phonon coupling strength $\lambda_{\rm ph}=0.4$. Note that the phonon alone  has a $T_c\sim 0.1$K. While we have been considering the effect of a phonon that does not hybridize with the plasmon, it has also been shown that polaritons formed by resonant coupling of longitudinal plasmons and phonons can enhance the $T_c$~\cite{Riolo2025}.
Alternatively, we may also consider the collaboration of the plasmon with spin-fluctuations which have been proposed as a mechanism for high-$T_c$~\cite{Scalapino1986, Scalapino2012}.

It should be noted that our treatment relies on the isotropic Eliashberg equations and neglects vertex corrections beyond Migdal's theorem, which may be quantitatively important when the plasmon energy is not small compared to the Fermi energy. A more complete treatment including anisotropic pairing channels and vertex corrections would sharpen the $T_c$ estimates.

\subsection{Candidate material platforms}\label{sec:materials}
Finally, we discuss potential material candidates that can support the model band structures examined in this work. Key requirements of a candidate material are that (i) there must be a coexistence of a light dispersive band and heavy nearly flat band, (ii) either the light or heavy band exhibits non-trivial quantum geometry, and (iii) experimental tunability of the band parameters. We identify multilayer moir\'e and dice lattices as groups of materials that can satisfy these requirements. For example, alternating-twist magic angle twisted trilayer graphene hosts a coexisting pair of flat bands and dispersive Dirac cones which are forbidden from hybridizing by mirror symmetry ~\cite{Khalaf2019, Lei2021, Park2021a, Hao2021, Pierce2025}. Alternating-twist multilayer graphene beyond three layers also exhibit similar characteristics, hosting a flat band pair and multiple dispersive bands ~\cite{Zhang2022, Park2022d}. Interband plasmons have also been observed in twisted bilayer graphene~\cite{Hesp2021}.  Such moir\'e systems are highly tunable through a variety of experimental knobs such as twist, stacking, and electrostatic gating, making them ideal material candidates for our work. Alternatively, the dice lattice has also been proposed to host a flat band that intersects with a dispersive set of bands~\cite{Sutherland1986} and has recently been experimentally observed in layered electride YCl~\cite{Geng2026}. These platforms therefore provide promising settings for combining heavy-electron plasmons, tunable band structures, and nontrivial quantum geometry to engineer superconductivity.

\section{Conclusion}
Our work establishes a flat-band mediator paradigm for superconductivity. Beyond their established role as hosts of superconductivity~\cite{Peotta2015,Liang2017,Hofmann2022,Hofmann2023}, flat-band subsystems can function as a tunable pairing mediator whose collective charge excitations generate retarded interactions in a separate itinerant sector. The superconducting scale is then governed by the collective-mode spectrum, coupling, and retardation rather than by the narrow single-particle bandwidth alone. This perspective shifts the design problem toward identifying flat bands that support well-defined collective modes at favorable energy scales while remaining strongly coupled to mobile carriers. More broadly, separating the subsystem that generates an interaction from the subsystem that develops order suggests a general strategy for engineering correlated phases through flat-band collective dynamics.

\emph{Note added.}---While finalizing this manuscript, we became aware of an independent work by Wang, Sarma, and Sau, whose preprint appeared recently on arXiv~\cite{Wang2026}. Their results partially overlap with ours in regarding intraband plasmons in metallic heavy electron bands. In particular, their observation of a dome-like behavior in the $T_c$ as a function of the heavy electron carrier density is in agreement with our results in \cref{sec:intraband}. The interband plasmon and quantum geometric screening effects are not discussed in their work~\cite{Wang2026}.

\begin{acknowledgments}
We thank Shuyang Wang, Sankar Das Sarma, and Jay Sau for informing us of their independent work on a related problem.
This work was supported by J.A.'s faculty start-up grant from The University of Texas at Austin.
\end{acknowledgments}

\FloatBarrier
\bibliography{references}
\end{document}